\documentclass[preprint,12pt]{elsarticle}

\usepackage[top=1in,bottom=1in,left=1in,right=1in]{geometry}

\usepackage{amssymb}
\newcommand{\Tr}{\mbox{Tr}}
\usepackage{amsthm, amsmath}

\usepackage{txfonts}
\usepackage{setspace}
\usepackage{xcolor}
\usepackage{caption}
\usepackage{subcaption}
\usepackage{float}
\usepackage[pdfborder={0 0 0},colorlinks,allcolors=blue]{hyperref}
\theoremstyle{definition}
\usepackage{comment}
\usepackage{xcolor, soul}
\usepackage{diagbox}
\usepackage{multirow}

\definecolor{fgwhite}{rgb}{1,1,1}     
\definecolor{fgred}{rgb}{0.8,0,0}     
\definecolor{fgorange}{rgb}{0.93,0.53,0.18}     
\definecolor{fgpurple}{rgb}{0.55,0.1,0.6}     
\definecolor{fggreen}{rgb}{0,0.5,0}     
\definecolor{bggreen}{rgb}{0.8,1,0.8}     
\definecolor{fgblue}{rgb}{0,0,0.7}     
\definecolor{bgblue}{rgb}{0.9,0.9,1}     
\definecolor{fgclay}{rgb}{0.51,0.25,0.04}     
\definecolor{bggreen}{rgb}{0.8,1,0.8}     

\allowdisplaybreaks

\journal{TBD}

\begin{document}

\begin{frontmatter}

\title{Structured force reformulation of many-body dispersion: towards effective atom--atom decomposition and surrogate modeling}
\author[inst1]{Zhaoxiang Shen}

\affiliation[inst1]{organization={Department of Engineering; Faculty of Science, Technology and Medicine; University of Luxembourg},
            city={Esch-sur-Alzette},
            postcode={4365}, 
            country={Luxembourg}}

\author[inst2]{Ra\'ul I. Sosa}
\author[inst1]{St\'ephane P.A. Bordas}
\author[inst2]{Alexandre Tkatchenko}
\author[inst1,inst3]{Jakub Lengiewicz}

\affiliation[inst2]{organization={Department of Physics and Materials Science, University of Luxembourg},
            city={Luxembourg City},
            postcode={1511}, 
            country={Luxembourg}}

\affiliation[inst3]{organization={Institute of Fundamental Technological Research, Polish Academy of Sciences},
            city={Warsaw},
            country={Poland}}

\begin{abstract}
We present a structured force reformulation of the many-body dispersion (MBD) model that enables a physically consistent decomposition of forces into pairwise components. By introducing a many-body correlation matrix that scales dipole--dipole interactions, we derive unified expressions for the MBD energy, force, and Hessian. This reformulation reveals a natural structure for effective atom--atom force decomposition and provides a promising foundation for interpretable analysis and machine learning surrogate modeling of MBD interactions.
\end{abstract}


\end{frontmatter}

\section{Introduction}
\label{sec:Introduction}
Van der Waals (vdW) dispersion interactions are essential for accurately modeling molecular systems, especially those governed by weak, long-range forces such as layered materials \cite{liu2019van}, molecular crystals \cite{ranking_crystal,stability_crystal}, and biomolecules \cite{ProteinFolding}. The many-body dispersion (MBD) model \cite{PhysRevLett.108.236402,doi:10.1063/1.4789814} represents a significant advance in this area by capturing quantum-mechanical electron correlation effects beyond the pairwise (PW) approximation. Unlike conventional PW functionals, MBD accounts for collective response behavior of atoms through coupled quantum harmonic oscillators, leading to forces that differ in both magnitude and direction compared to their pairwise counterparts. These many-body corrections have been shown to yield improved agreement with experimental observations \cite{Polymorphism,polymorphs,Supramolecular,Hauseux_NC,OrganicMolecularMaterials} and reveal non-trivial behavior in a wide range of systems \cite{Stohr_water,Colossal_hessian,PRR_Mario_23, wavelike_science_2016,ambrosetti_optical}.

However, the MBD formalism involves the diagonalization of a global dipole–dipole coupling matrix, resulting in an $O(N^3)$ computational cost and obscuring intuitive atomic-level interpretation. In contrast to simple PW models, where forces can be naturally attributed to atomic pairs, the collective nature of MBD makes such decompositions non-trivial. This limits not only the interpretability of MBD results but also the development of efficient machine learning (ML) surrogate models, known as machine learning force fields (MLFFs) \cite{MLFF-2020,MLFF-2021-1,MLFF-2021-2,MLFF-2021-3,MLFF-book}, to accelerate its application. A tool that enables atomic- or pairwise-level decomposition of MBD interactions would thus be considerably valuable for both analysis and acceleration.

Several previous efforts have attempted to partition MBD quantities to allow physically insightful analysis and practical computation. For example, a second quantization \cite{book_SQ} approach by Gori et al. \cite{2ndQuantMBD} partitions the MBD energy across molecular fragments, while a method proposed by Galante et al. \cite{PRR_Mario_23} decomposes the MBD force according to the displacement vectors of interacting dipoles. These efforts offer valuable insight into the nature of many-body correlations while preserving energy or force consistency. Another approach \cite{atom-wise_MBD} explores atom-wise contributions with computational approximations to speed up large-scale MBD calculations, but lacks an exact and consistent physical decomposition. Despite these efforts, a generic, physically grounded, and computationally practical decomposition framework remains elusive, especially one that is beneficial for ML surrogate modeling.

In this work, we propose a structured force reformulation of the MBD model that enables a physically consistent and interpretable decomposition of the force. Central to our approach is the introduction of a matrix $\boldsymbol{B}$, referred to as the \emph{many-body correlation factor}, which encodes system’s many-body correlations and scales the pairwise dipole–-dipole coupling matrix $\boldsymbol{C}$. This leads to a unified reformulation of the MBD with clean expressions for the energy, force, and Hessian, expressed entirely in terms of $\boldsymbol{B}$ and $\boldsymbol{C}$ or its derivatives. The reformulation reveals a natural and structured decomposition of the force into pairwise-like components of the form $b\nabla c$. We demonstrate this framework using low-dimensional model systems: parallel carbon chains and rings, which highlight the characteristic behavior and interpretability of the reformulated and decomposed MBD expressions.

Finally, this framework provides a promising foundation for ML surrogate modeling of MBD interactions. Despite the growing interest in MLFFs, no general solution currently exists for explicitly representing MBD interactions within a surrogate learning framework. The structured nature of $\boldsymbol{B}$ offers a novel and expressive representation for the learning task, enabling new surrogate modeling strategies based on the decomposition. To support future development, we open-source the data generation codes \cite{MBD_reform_Repo} capable of efficiently generating the training data based on the MBD reformulation.

\section{Many-body dispersion: theory and preliminary analysis}
\subsection{Fundamental MBD formulation}
\label{sec:mbd}
Van der Waals (vdW) dispersion, also known as London dispersion, is a ubiquitous attractive interaction among atoms and molecules, originating from temporary mean-field electron density fluctuations that create instantaneous dipoles. Traditional models compute the dispersion energy for an $N$-atom molecular system via PW interactions:

\begin{equation}
E^{\text{PW}} = -\sum_{j>i}^{N}f^\text{damp} \dfrac{\mathcal{C}_{6,ij}}{r_{ij}^{6}},
\label{eq:E_pw}
\end{equation}
where $r_{ij}=\|\boldsymbol{r}_{i}-\boldsymbol{r}_{j}\|$ is the interatomic distance between atoms $i$ and $j$, and $\boldsymbol{r}$ represents the atomic position. The material parameters $\mathcal{C}_{6,ij}$ are determined experimentally or through high-fidelity numerical calculations. A damping function $f^\text{damp}$ is employed to mitigate short-range singularities. The PW force of an atom $i$ in the system is given by: 
\begin{equation}
\boldsymbol{F}_{i}^{\text{PW}}=-\nabla_{i}E^{\text{PW}}.
\end{equation}

However, vdW dispersion is intrinsically many-body in nature due to correlated electron fluctuations. The many-body dispersion (MBD) method treats the electronic response via atom-centered quantum harmonic oscillators (QHOs) coupled by a dipolar potential \cite{doi:10.1063/1.4789814, ambrosetti2014long}. In this framework, the dipole-dipole coupling tensor $\boldsymbol{C}$ is composed of $N^2$ of $3\times3$ blocks that describe the coupling between each pair of atoms $i$ and $j$:
\begin{equation}
    \boldsymbol{C}_{ij}^{\text{}} = \omega_i^2 \delta_{ij} + (1-\delta_{ij}) \ \omega_i\omega_j \sqrt{ \alpha_{i}^\text{0,eff} \alpha_{j}^\text{0,eff}}\ \boldsymbol{T}_{ij}.
    \label{eq:Cmat}
\end{equation}
where $\omega_i$ and $\alpha^\text{0,eff}$ are the characteristic frequency and the effective polarizability for atom $i$, and $\boldsymbol{T}_{ij}$ represents the dipole-dipole interaction for two QHOs at atoms $i$ and $j$. The MBD energy is then obtained from the ground-state energy shift of the coupled QHO system:
\begin{equation}
E^{\text{MBD}} = \dfrac{1}{2} \Tr(\boldsymbol{\Lambda}^{\frac{1}{2}}) - \dfrac{3}{2} \sum_{i=1}^{N} \omega_i, 
\label{eq:E_mbd}
\end{equation}
where the diagonal matrix $\boldsymbol{\Lambda}_{pq}=\lambda_p\delta_{pq}$ is obtained from diagonalizing $\boldsymbol{C}$ via an orthogonal transformation matrix $\boldsymbol{S}$ such that $\boldsymbol{\Lambda}=\boldsymbol{S}^\text{T}\boldsymbol{C}\boldsymbol{S}$. The first term of the energy accounts for the collective mode frequencies, and the second term corresponds to the non-interacting frequencies. Taking the negative gradient of the MBD energy with respect to atomic positions leads to the MBD force at atom $i$: 
\begin{equation}
\boldsymbol{F}_{i}^{\text{MBD}}= -\frac{1}{4} \Tr \left[ \boldsymbol{\Lambda}^{-\frac{1}{2}}\, \boldsymbol{S}^\text{T} \left({\nabla}_{i}\boldsymbol{C}\right)\boldsymbol{S}\right],
\label{eq:F_mbd}
\end{equation}
which follows the standard differentiation rules for eigenvalues \cite{Diff_eigen}.

\subsection{MBD forces on low-dimensional systems}
\begin{figure}[h]
 \centering
    \subfloat[Parallel carbon chains.]{\includegraphics[trim = -5mm -15mm 90mm 0mm, clip=true,width=0.30\textwidth]{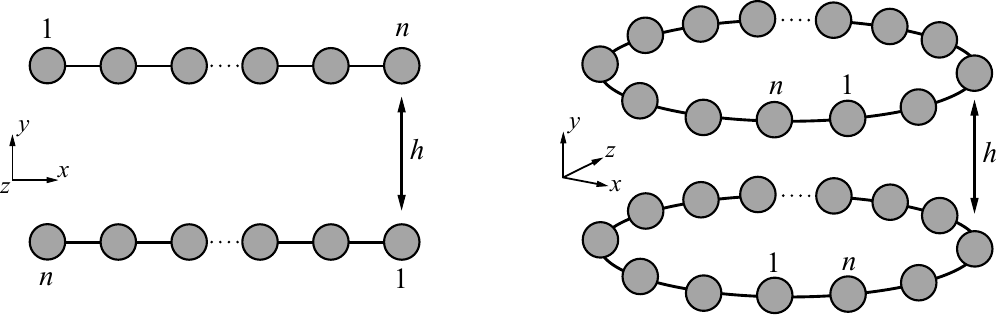}\label{fig:Geo_chain}}\hfil
    \subfloat[Parallel carbon rings.]{\includegraphics[trim = 90mm -15mm -5mm 0mm, clip=true,width=0.30\textwidth]{figures/chain_ring_diagram.pdf}\label{fig:Geo_ring}}\hfil
    \subfloat[$F_y$ over the fragment of upper chain or ring.]{\includegraphics[trim = 30mm 90mm 40mm 90mm, clip=true,width=0.36\textwidth]{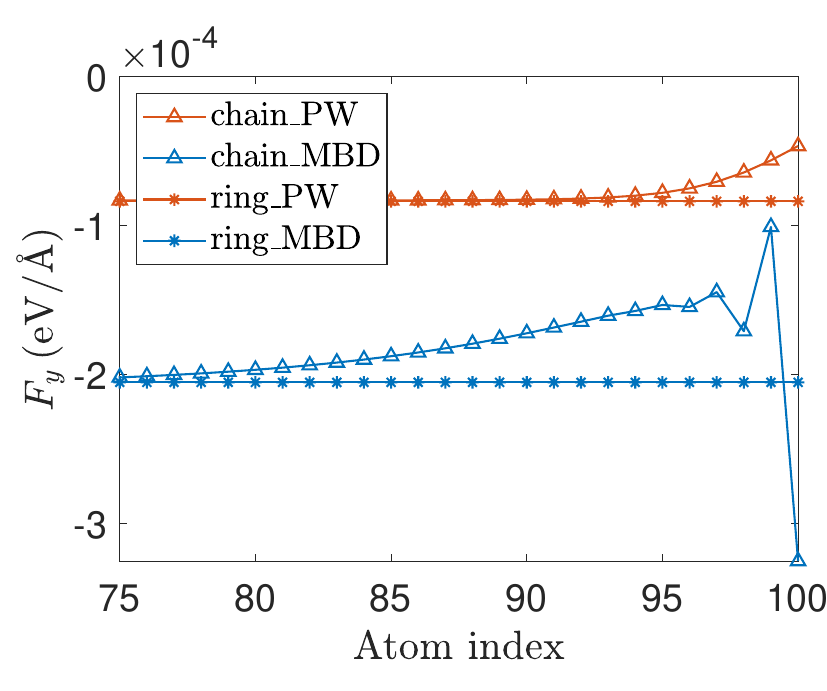}\label{fig:Fmbd_Fpw_chain_ring}}
    \caption{(a) and (b) show schematic representations of two parallel carbon chains and two parallel carbon rings, respectively, separated by a distance $h$ along the $y$ direction. Each molecule contains $n$ atoms ($N=2n$) with a uniform interatomic spacing of $1.2\,\text{\AA}$, and atoms in the two chains or rings are perfectly aligned along the $y$ axis. These are fixed, fictitious systems, with no consideration of covalent bonding. Atoms in the upper and lower structures have reverse numbering to facilitate heatmap analysis in subsequent sections. (c) shows the $F_y$ force profile along the upper chain or ring computed using different vdW dispersion models for systems with $n=100$ and $h=10\,\text{\AA}$.}\label{fig:chain_ring_geo_force}
    
\end{figure}

The diagonalization required by the MBD formulation not only introduces an $O(N^3)$ computational cost but also reveals the collective nature of MBD interactions, complicating straightforward interpretation and atomic-level analysis. Previous insightful analyses connecting these interactions directly to geometric features have primarily considered simplified, low-dimensional molecular structures~\cite{DFTB+MBD,Colossal_hessian,Hauseux_NC,wavelike_science_2016}, where dispersion interactions predominantly occur along a single spatial direction. However, such analyses typically focus on global responses or rely on Hessian-based approaches to examine pairwise interactions. To facilitate the development of a practical force decomposition for MBD, it is therefore beneficial to first explicitly investigate these interactions at the atomic force level in simplified molecular systems, as illustrated in Fig.~\ref{fig:chain_ring_geo_force}.

Here, we focus on elementary carbon-based wire structures (chains and rings), analyzing vdW dispersion interactions between two parallel and perfectly aligned chains (or rings). We compare two vdW models: the Tkatchenko–Scheffler PW model \cite{PhysRevLett102073005} and the plain MBD model \cite{DFTB+MBD}. As shown in Fig.~\ref{fig:Fmbd_Fpw_chain_ring}, the PW model clearly reflects the expected asymmetric behavior at the chain edges, while the MBD model additionally exhibits a non-trivial wavy force profile. Motivated by this anomalous observation, we introduce rings as our second test system, in which the edges are connected to remove boundary asymmetry, resulting in a uniform force response across the structure for both PW and MBD interactions. 

The intriguing force profiles observed in both the chain and ring systems, where symmetry governs the nature and balance of interactions, suggest that even in a many-body setting, the geometric features of the system play a decisive role. These observations encourage the development of a geometrically consistent approach for MBD force decomposition, one that can account for and explain the subtle spatial variations in the force profiles. In the next section, we introduce our reformulated MBD framework, incorporating a dedicated many-body correlation factor, as the basis for a structured atom--atom decomposition. This formulation provides a transparent way to rationalize the observed phenomena and serves as a reference point for further advancements in surrogate modeling of MBD interactions.

\section{Reformulation of the MBD formulation}
\label{sec:MBD_reform}
As shown in the previous section, computing the MBD energy or force requires diagonalization of $\boldsymbol{C}$, followed by application of the trace operator to the resulting eigendecomposition products. While this process involves a cubic-scaling computational cost, the intrinsic structure of the MBD formulas permits flexible algebraic rearrangements due to the commutative properties of the trace operator and the orthogonality of the eigenvector matrix $\boldsymbol{S}$. This flexibility allows for reformulating the MBD formulation within a purely algebraic framework. Motivated by the goal of enabling pairwise analysis of MBD, we seek a new construction that explicitly preserves pairwise quantities. For this reason, we introduce the ``many-body correlation factor'':
\begin{equation}
    \boldsymbol{B}=\boldsymbol{S}\boldsymbol{\Lambda}^{-\frac{1}{2}}\boldsymbol{S}^T=\boldsymbol{C}^{-\frac{1}{2}},
\end{equation}
which helps to decouple the many-body and pairwise components embedded in the original formulation. We start reformulation from the energy formula given by Eq.~\eqref{eq:E_mbd}:
\begin{equation}
\begin{aligned}
E^{\text{MBD}} &= \dfrac{1}{2} \Tr{\left ( \boldsymbol{\Lambda}^{\frac{1}{2}}\right )} - \dfrac{3}{2} \sum_{i=1}^{N} \omega_i \\
&= \dfrac{1}{2} \Tr{\left ( \boldsymbol{C}^{\frac{1}{2}}\right )} - \dfrac{3}{2} \sum_{i=1}^{N} \omega_i \\
&= \dfrac{1}{2} \Tr{\left (\boldsymbol{B}\boldsymbol{C}\right )} - \dfrac{3}{2} \sum_{i=1}^{N} \omega_i,
\end{aligned}
\label{eq:E_mbd_reform}
\end{equation}
which leverages the similarity invariance of the trace. While the trace of $\boldsymbol{B}\boldsymbol{C}$ remains mathematically equivalent to the summation over interacting QHO mode frequencies, it now takes the form of a pairwise coupling matrix scaled by a many-body correlation factor. This structure paves the way for a atom--atom decomposition of the MBD energy and potentially a unified reformulation of its gradient.

Building on the reformulated energy expression, we derive the MBD force in terms of $\boldsymbol{B}$:
\begin{equation}
    \begin{aligned}
            \boldsymbol{F}_{i}^\text{MBD} = -{\nabla}_{i}{E^{\text{MBD}}} &= -\frac{1}{2}\Tr\left(\boldsymbol{B}\left({\nabla}_{i}\boldsymbol{C}\right)+\left({\nabla}_{i}\boldsymbol{B}\right)\boldsymbol{C}\right) \\
        &=-\frac{1}{2}\Tr\left(\boldsymbol{B}\left({\nabla}_{i}\boldsymbol{C}\right)-\frac{1}2{}\boldsymbol{B}\left({\nabla}_{i}\boldsymbol{C}\right)\right)  \\
        &= -\frac{1}{4} \Tr \left(\boldsymbol{B}\left({\nabla}_{i}\boldsymbol{C}\right)\right),
    \end{aligned}
    \label{eq:F_mbd_reform}
\end{equation}
where the equivalence $\Tr\left({\boldsymbol{B}\left({\nabla}\boldsymbol{C}\right)}\right)=-2\Tr\left(\left({\nabla}\boldsymbol{B}\right)\boldsymbol{C}\right)$ is proved in \ref{sec:app_gradB}. The original force expression in Eq.~\eqref{eq:F_mbd} can be recovered from the reformulated version using the cyclic property of the trace. Importantly, this new expression preserves the linear structure of the interacting energy from Eq.~\eqref{eq:E_mbd_reform}, with differentiation acting only on the pairwise term $\boldsymbol{C}$. This highlights the consistency and conceptual clarity of the proposed reformulation. Compared with the energy formula, the reformulated force expression becomes a more effective and physically insightful tool for pairwise analysis of MBD. Forces, as vector quantities, naturally localize interactions through their directional dependencies, enabling clearer attribution of physical effects to individual atom pairs. Moreover, due to the pairwise construction of $\boldsymbol{C}$, its gradient $\nabla_i\boldsymbol{C}$ only retains dipole--dipole couplings involving the atom $i$, while the global many-body correlation is fully encoded in $\boldsymbol{B}$. For these reasons, our MBD decomposition will be developed only at the force level with Eq.~\eqref{eq:F_mbd_reform}, see details in the next section.

\begin{figure}[htbp]
 \centering

    \subfloat[Chain, condensed $\boldsymbol{B}$.]{\includegraphics[trim = 0mm 60mm 0mm 60mm, clip=true,width=0.49\textwidth]{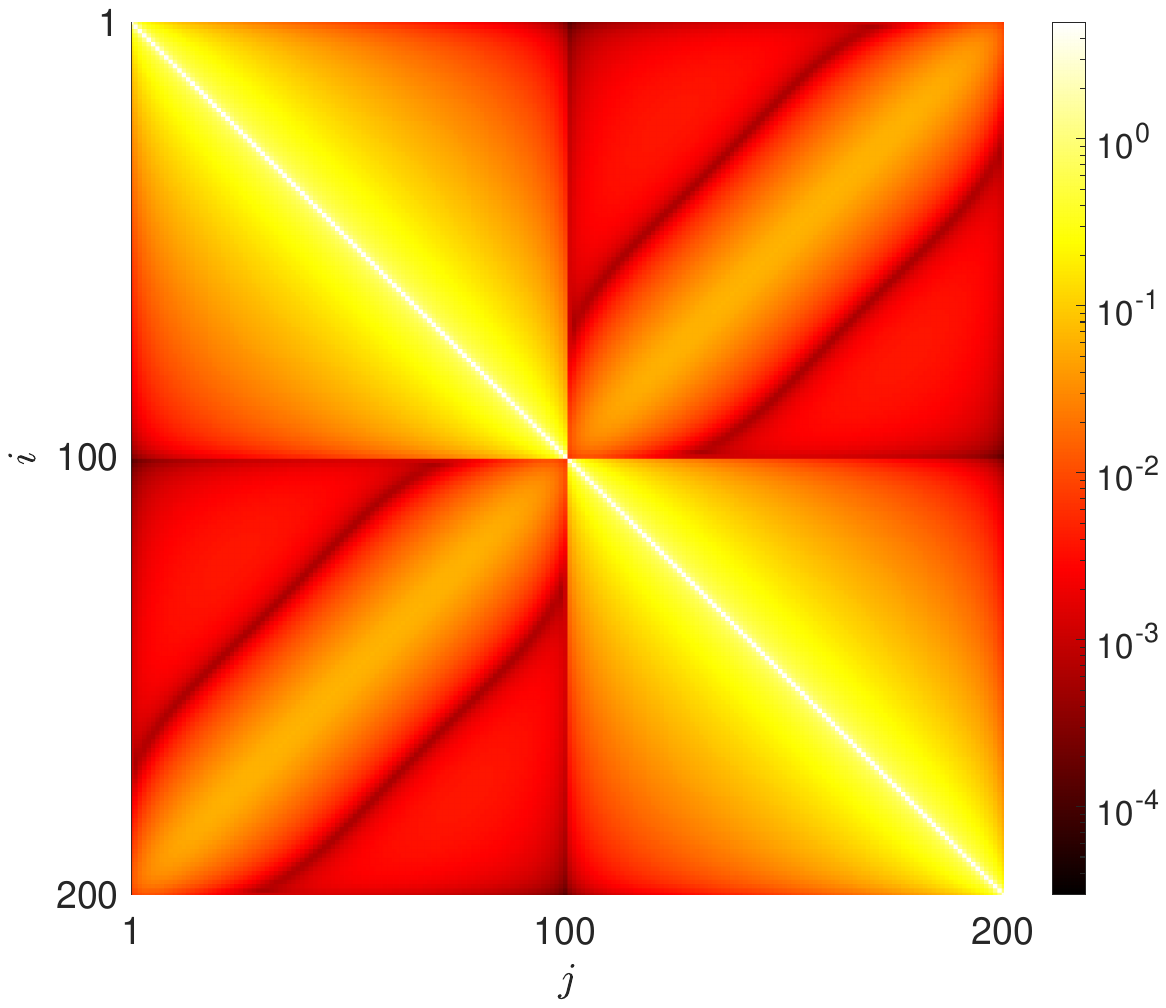}\label{fig:Fmbd_chain_conB}}
    \subfloat[Chain, condensed $\nabla\boldsymbol{C}$.]{\includegraphics[trim = 0mm 60mm 0mm 60mm, clip=true,width=0.49\textwidth]{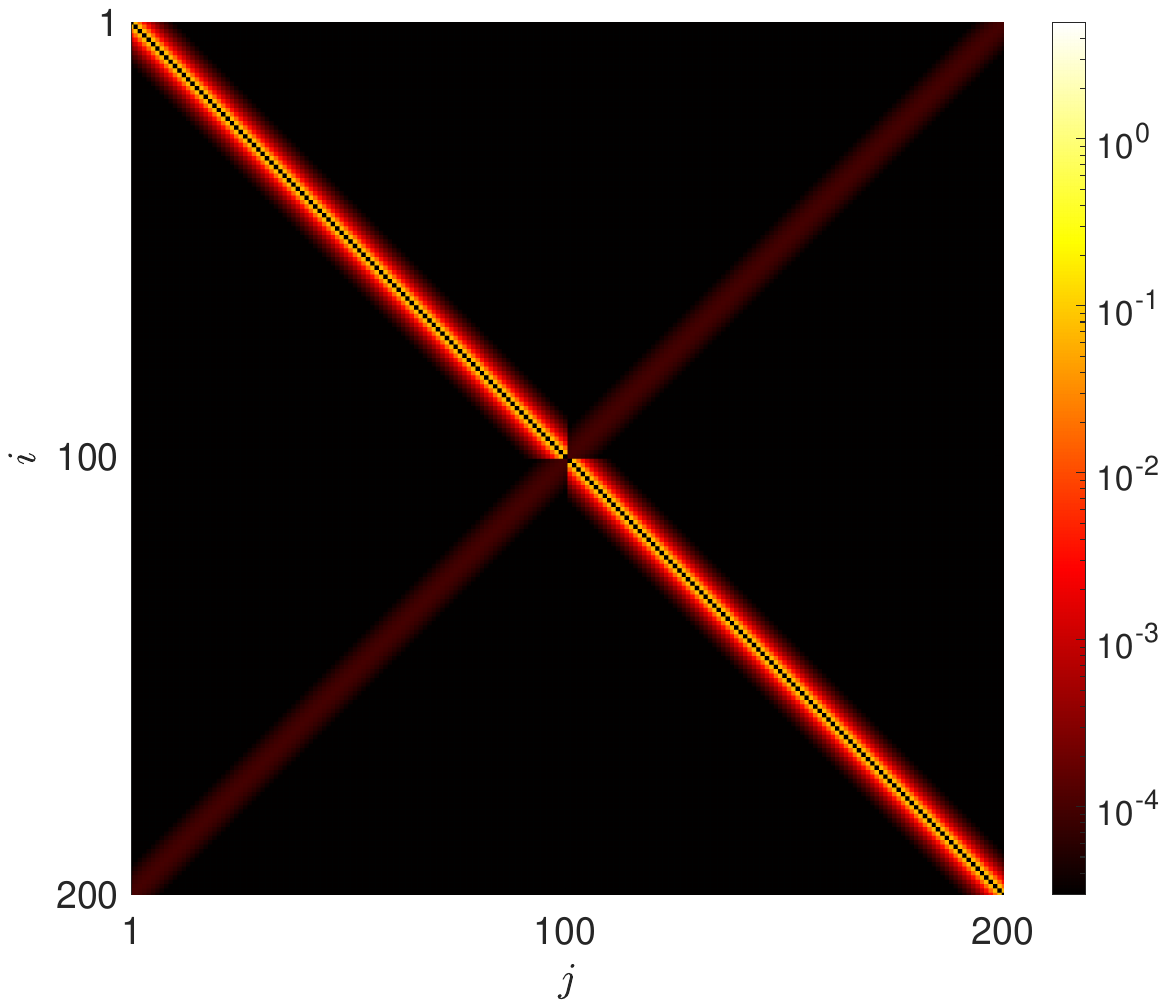}}\hfill
    \subfloat[Ring, condensed $\boldsymbol{B}$.]{\includegraphics[trim = 0mm 60mm 0mm 60mm, clip=true,width=0.49\textwidth]{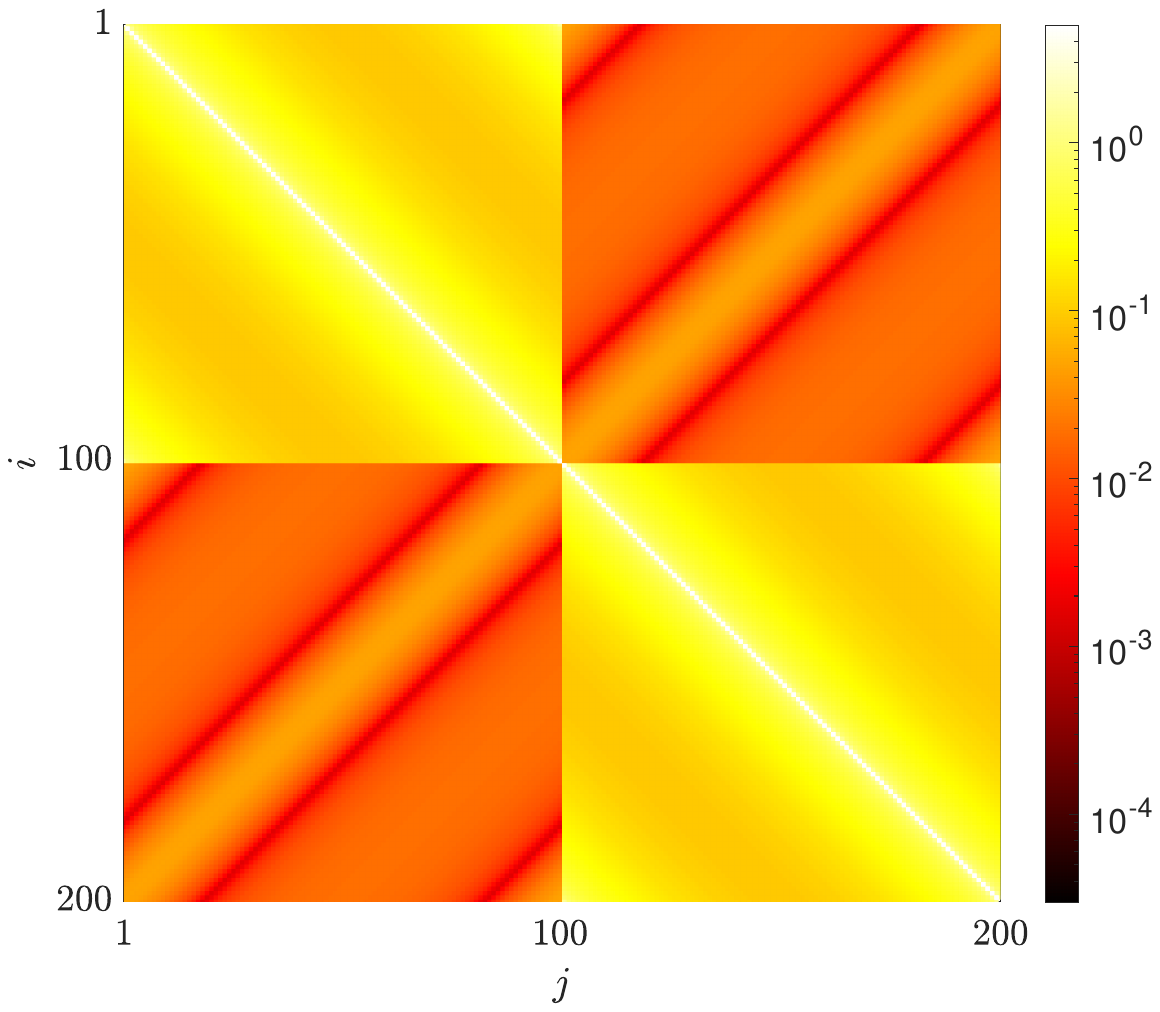}\label{fig:Fmbd_ring_conB}}
    \subfloat[Ring, condensed $\nabla\boldsymbol{C}$.]{\includegraphics[trim = 0mm 60mm 0mm 60mm, clip=true,width=0.49\textwidth]{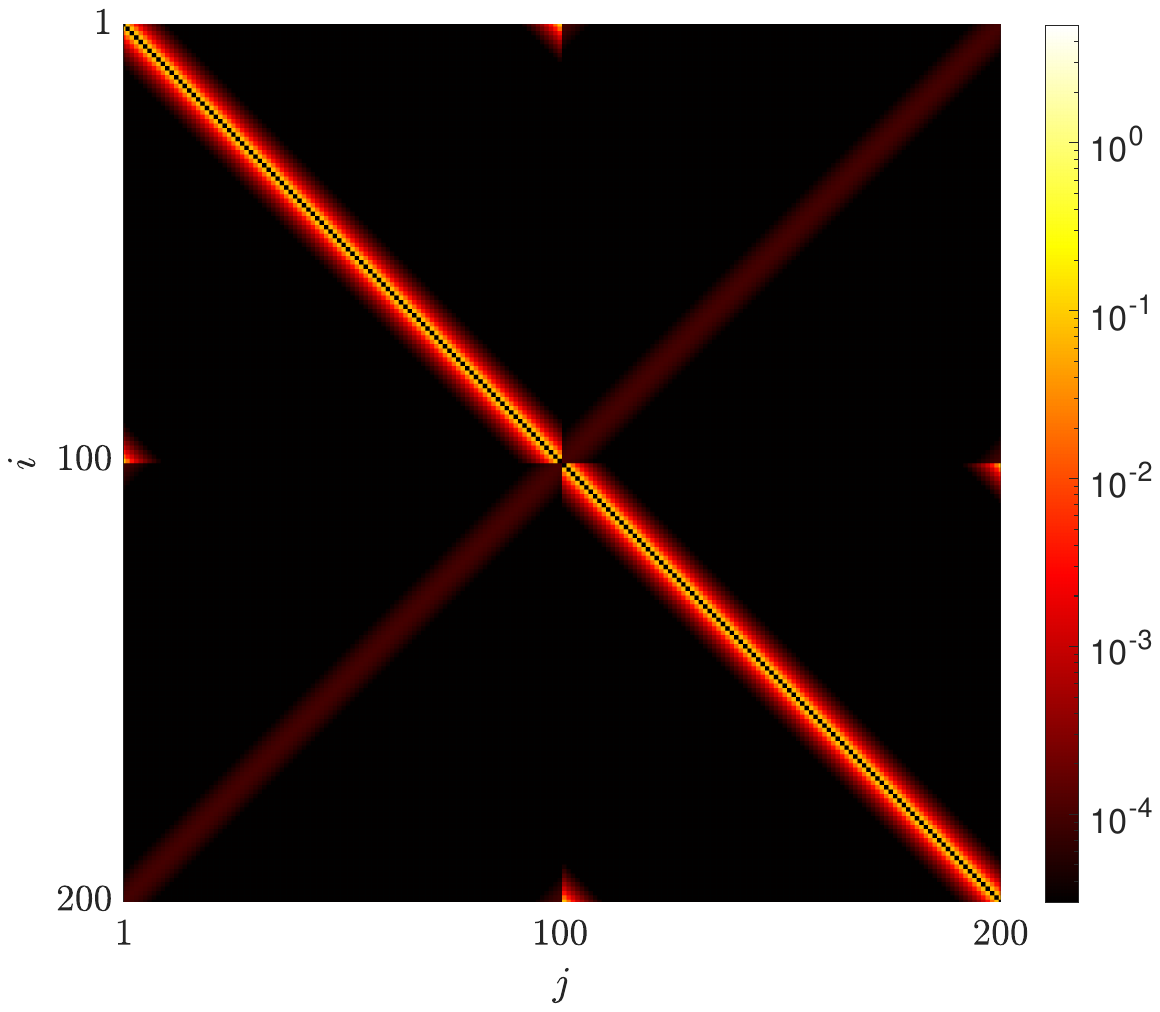}}
    \caption{Heatmaps of condensed $\boldsymbol{B}$ and $\nabla\boldsymbol{C}$, see definition in Eq.~\eqref{eq:condensed_mat}, for the two molecular systems (chain and ring) with $n=100$, and $h=10\,\text{\AA}$. The plot axes correspond to atomic indices $i$ and $j$, where the first 100 indices represent atoms in the upper molecule and the second 100 represent atoms in the lower molecule, ordered in reverse (see Fig.~\ref{fig:chain_ring_geo_force}). Under this convention, the main diagonal blocks correspond to intra-molecular interactions, while the off-diagonal blocks capture inter-molecular interactions.}
    \label{fig:Fmbd_chain_ring_conBdC}

\end{figure}
To further illustrate the structural properties of $\boldsymbol{B}$ and $\nabla \boldsymbol{C}$, we analyze their matrix structures for the simple molecular systems presented in Fig.~\ref{fig:chain_ring_geo_force}: parallel carbon chains (Fig.~\ref{fig:Geo_chain}) and parallel carbon rings (Fig.~\ref{fig:Geo_ring}). To capture the essential physical characteristics of each matrix, we depict their condensed versions in Fig.~\ref{fig:Fmbd_chain_ring_conBdC}, in which each pairwise $3\times3$ block is replaced by its Frobenius norm. Specifically, we define the condensed matrices as follows:
\begin{equation}
    b_{ij}^\text{cond} = \sqrt{\sum_{\alpha,\beta}^{x,y,z}\left(b_{i,\alpha,j,\beta}\right)^2}, \quad\quad    \left(\nabla_{i,y} c\right)_{ij}^\text{cond} = \sqrt{\sum_{\alpha,\beta}^{x,y,z}\left(\nabla_{i,y}c_{i,\alpha,j,\beta}\right)^2},
    \label{eq:condensed_mat}
\end{equation}
where the gradient is taken only in the $y$ direction, which is the dominant interaction axis in the selected systems. The heatmaps for the condensed $\boldsymbol{B}$ illustrate its many-body nature through pronounced non-trivial contributions from long-range interactions across the entire molecular structure. In contrast, the condensed $\nabla\boldsymbol{C}$ exhibits strong locality, consistent with the short-range, pairwise nature of dipole--dipole coupling. Moreover, the condensed $\boldsymbol{B}$ reveals the geometrical characteristics of the molecular systems. For the ring structure, atoms in each ring experience identical local atomic environments, resulting in a regular and symmetric heatmap pattern, as shown in Fig.~\ref{fig:Fmbd_ring_conB}. On the other hand, the heatmap of the chain structure is influenced by the presence of asymmetric side atoms, leading to a slightly irregular pattern, see Fig.~\ref{fig:Fmbd_chain_conB}. The most intriguing feature is the appearance of a wavy off-diagonal profile, representing interactions between opposite sides of the chain or ring structures. This observation is particularly notable, as it may be related to the wavelike behavior of MBD interactions reported in \cite{wavelike_science_2016,Hauseux_NC}. We will further discuss this phenomenon in detail during the pairwise decomposition analysis presented in the following section.

Lastly, we note that the reformulation also permits an analytical expression for the MBD Hessian:
\begin{equation}
    \begin{aligned}
            \boldsymbol{H}_{ij}^\text{MBD} = {\nabla}_{i}{\nabla}_{j}{E^{\text{MBD}}} &= \frac{1}{4}\Tr\left(\boldsymbol{B}\left({\nabla}_{i}{\nabla}_{j}\boldsymbol{C}\right)+\left({\nabla}_{i}\boldsymbol{B}\right)\left({\nabla}_{j}\boldsymbol{C}\right)\right).
    \end{aligned}
    \label{eq:hessian}
\end{equation}
The explicit expression for $\nabla\boldsymbol{B}$ is involved and cannot easily be represented in a concise matrix form (see details in \ref{sec:app_gradB}). Nevertheless, this Hessian formula still preserves the linear structure with ``MBD$\times$PW'', as the additional gradient does not alter the inherent physical nature of $\boldsymbol{B}$ and $\boldsymbol{C}$. Given the non-locality of $\boldsymbol{B}$ illustrated in Fig.~\ref{fig:Bmat_ML_diagram}, the obtained analytical formula offers a systematic explanation for the pronounced long-range behavior observed in MBD Hessian heatmap presented in \cite{Colossal_hessian}. While an in-depth analysis of the Hessian is beyond the scope of the current study, we will briefly highlight its computational utility in Section~\ref{sec:MLMBD}, as it offers the potential for efficient and exact Hessian evaluation.

\section{Decomposition of atomic MBD force}
As discussed in the previous section, the reformulated MBD force formula effectively isolates the many-body information in $\boldsymbol{B}$ and pairwise description in $\nabla\boldsymbol{C}$. This linear structure naturally motivates an effective atom--atom decomposition of the MBD force in a pairwise manner, beginning by rewriting Eq.~\eqref{eq:F_mbd_reform} in index notation:
\begin{equation}
    f_{i,\alpha} = -\frac{1}{2}b_{i,\gamma,j,\beta}\nabla_{i,\alpha} c_{i,\gamma,j,\beta},
\end{equation}
where $f_{i,\alpha}$, $b_{i,\gamma,j,\beta}$, and $c_{i,\gamma,j,\beta}$ denote the components of $\boldsymbol{F}_i^{\rm MBD}$, $\boldsymbol{B}$, and $\nabla\boldsymbol{C}$, respectively. Summation over $j$, $\beta$, $\gamma$ is silently assumed, where $j=1,\ldots, N$ and $\alpha, \beta, \gamma, \in \{x,y,z\}$. Here we utilize the pairwise nature of $\boldsymbol{C}$, therefore $\nabla\boldsymbol{C}$ has non-zero entries only in the off-diagonal $3\times3$ blocks associated with atom $i$. This expression is analogous to the standard PW formula for vdW interactions, where the force is expressed as a sum of contributions from all other atoms. In our case, this can be written as:
\begin{equation}
    f_{i,\alpha,j,\beta} = -\frac{1}{2}b_{i,\gamma,j,\beta}\nabla_{i,\alpha} c_{i,\gamma,j,\beta},
\end{equation}
where the summation over $\gamma$ is still assumed. One can alternatively derive this form directly from Eq.~\eqref{eq:F_mbd}, using the relation $b_{i,\gamma,j,\beta} = \sum_{p=1}^{3N}s_p^{i,\gamma}\lambda_p^{-\frac{1}{2}}s_p^{j,\beta}$. In the context of our structured MBD reformulation, this decomposition formally represents the interaction between atoms $i$ and $j$ as a many-body-scaled dipole-–dipole ``force''. Notably, it exhibits zero self-contribution ($f_{i,\gamma,i,\beta}=0$) due to the vanishing diagonal of $\nabla\boldsymbol{C}$, and preserves force symmetry between atoms ($f_{i,\gamma,j,\beta}=-f_{j,\beta,i,\gamma}$), which arises from the symmetry of $\boldsymbol{B}$ and the antisymmetry of $\nabla\boldsymbol{C}$. These properties give the expression a structured and physically consistent form that closely resembles a conventional PW force formulation.

To demonstrate the utility of the proposed force decomposition, we once again apply it to the same chain and ring systems discussed earlier. As the dominant interaction occurs along the $y$ direction, we visualize only the $f_{i,y,j}$ components, i.e., the contribution of each atom $j$ to the $y$ component of the force on atom $i$. For clarity, we aggregate the contributions from all three components of $j$ ($f_{i,y,j}=\sum_{\beta=x}^{z}f_{i,y,j,\beta}$), since attributing force contributions from a single DOF of $j$ to a specific DOF of $i$ is less physically meaningful. Importantly, these heatmaps retain the sign of the force contributions, capturing their physical directionality.

\begin{figure}[htbp]
 \centering
    \subfloat[Chain.]{\includegraphics[trim = 0mm 60mm 0mm 60mm, clip=true,width=0.33\textwidth]{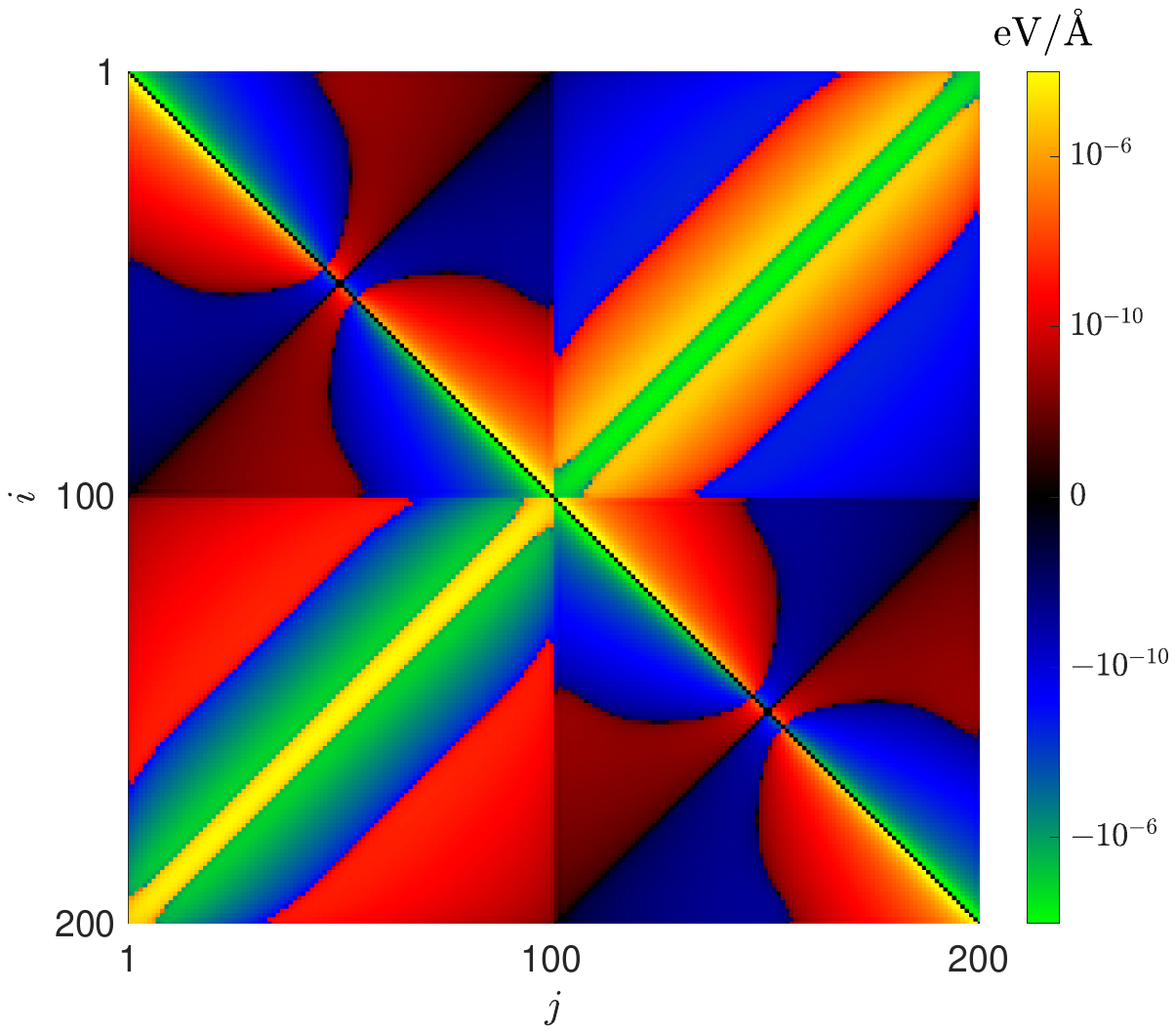}\label{fig:Fmbd_decompose_chain}}
    \subfloat[Ring.]{\includegraphics[trim = 0mm 60mm 0mm 60mm, clip=true,width=0.33\textwidth]{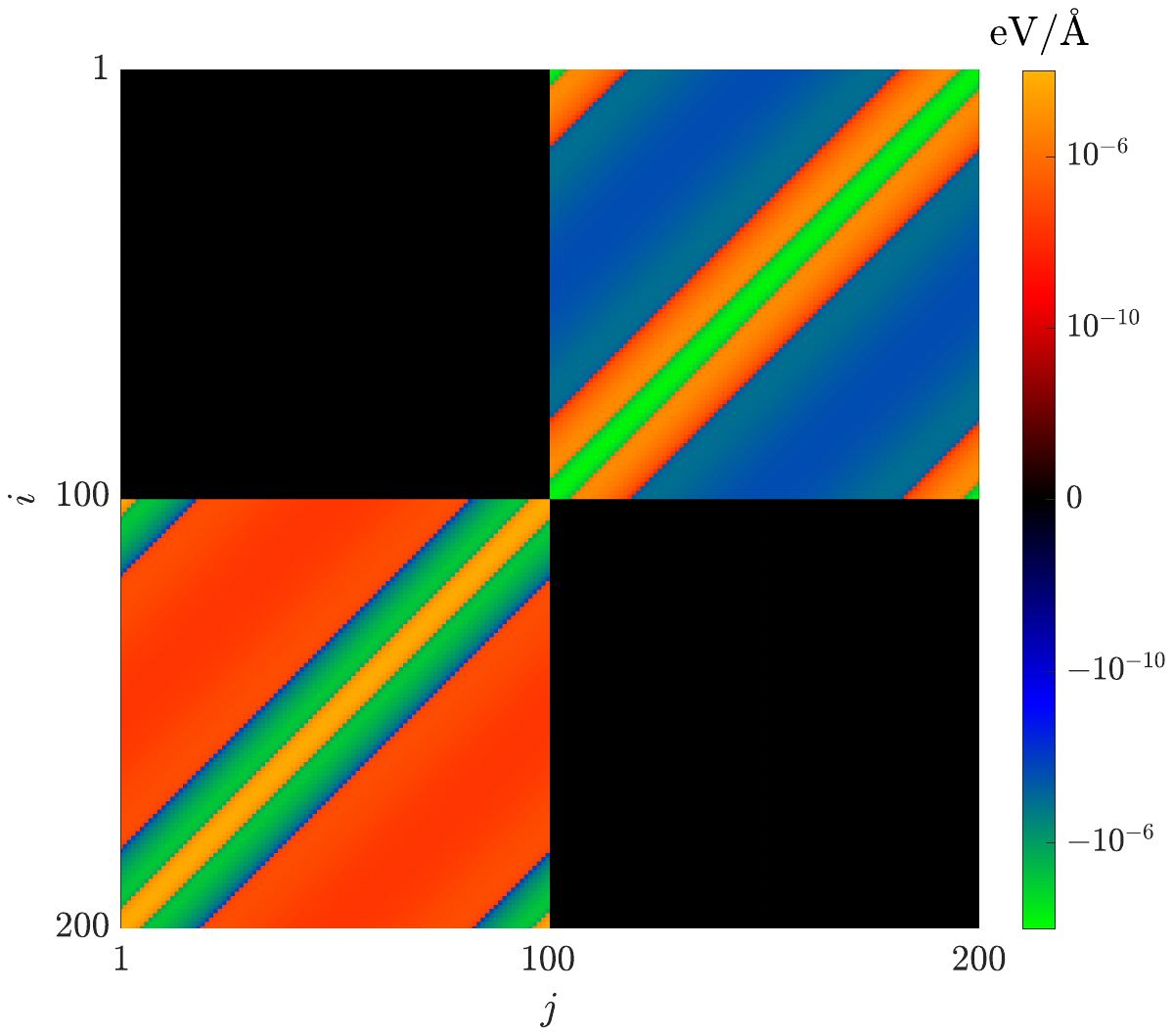}\label{fig:Fmbd_decompose_ring}}
    \subfloat[Ring with 1-atom vacancy.]{\includegraphics[trim = 0mm 60mm 0mm 60mm, clip=true,width=0.33\textwidth]{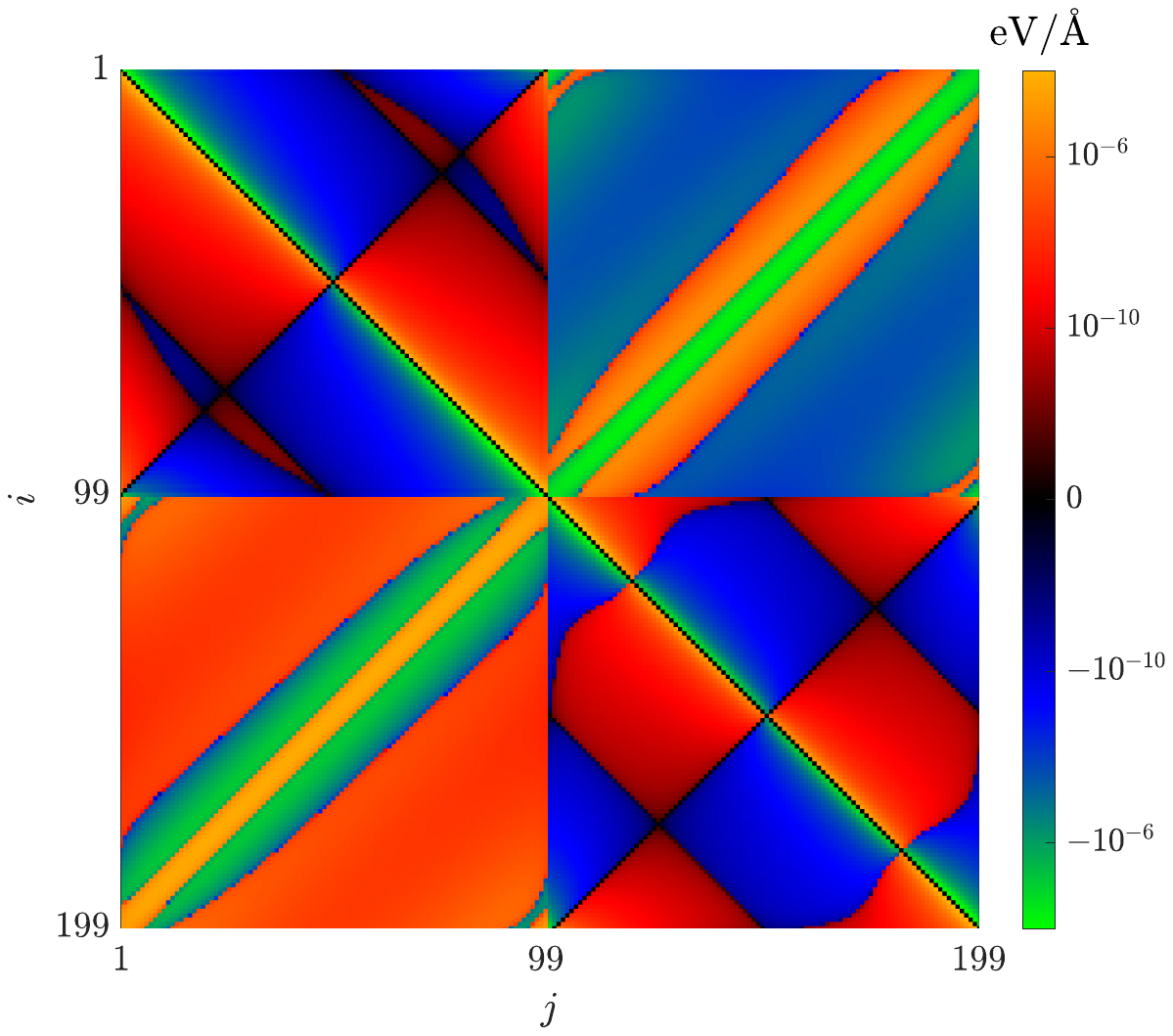}\label{fig:Fmbd_decompose_ring_vac}}

    \caption{Heatmaps of the MBD force decomposition component $f_{i,y,j}$ for the two molecular systems (chain and ring) with $h=10\,\text{\AA}$. Each value represents the contribution from atom $j$ to $f_{y}$ of atom $i$. The same atomic indexing convention is used as in Fig.~\ref{fig:Fmbd_chain_ring_conBdC}. In (a) and (b), the systems preserve symmetry between the upper and lower molecules with $n=100$. In (c), the last atom of the upper ring is removed ($n_\text{upper}=99$), while the lower ring remains unchanged.}
    \label{fig:Fmbd_decompose_chain_ring}

\end{figure}

From the heatmaps depicted in Fig.~\ref{fig:Fmbd_decompose_chain_ring}, we clearly see that the dominant contribution to the force component $f_{i,y}$ originates from atoms in the closest positions of the opposite chain or ring, which aligns intuitively with the parallel and flat arrangement of the molecular systems. The corresponding off-diagonal profiles closely resemble the wavelike patterns observed previously in the condensed $\boldsymbol{B}$ (Figs.~\ref{fig:Fmbd_chain_conB} and \ref{fig:Fmbd_ring_conB}) for each structure, apart from the alternating signs of individual atomic contributions. It is primarily this alternating sign behavior that leads to the characteristic wavy pattern, visible not only in the inter-chain/ring regions (off-diagonal blocks) but also apparently within intra-chain interactions (diagonal blocks) shown in Fig.~\ref{fig:Fmbd_decompose_chain} for the chain system.

A particularly counterintuitive observation is that neighboring atoms within the same chain, despite being aligned along the $x$ direction, contribute significantly to the perpendicular force component $f_{i,y}$. The net force of each atom ultimately results from a delicate balance of these opposing contributions. For atoms near the edges of the chain, the symmetry breaks down, explaining the irregular wavelike force distribution previously illustrated in Fig.~\ref{fig:Fmbd_Fpw_chain_ring}. Conversely, atoms located closer to the center of the chain experience a more symmetric environment, causing the wavy pattern from both sides to partially offset and smooth out.

This symmetry-induced cancellation is further confirmed by the decomposition heatmap for the ring systems, as shown in Fig.~\ref{fig:Fmbd_decompose_ring}. Here, the main diagonal blocks vanish, indicating a complete cancellation of intra-ring contributions. This result is consistent with the physical intuition of a perfectly closed ring structure, where each atom experiences an identical and perfect symmetric environment, thus aligning with the uniform force profile shown in Fig.~\ref{fig:Fmbd_Fpw_chain_ring}. Consequently, unlike the chain system, the forces within the ring arise exclusively from inter-ring contributions. However, when the perfect symmetry is disrupted by removing a single atom from the upper ring, the characteristic wavy pattern immediately reappears within the diagonal block (see Fig.~\ref{fig:Fmbd_decompose_ring_vac}). Simultaneously, the off-diagonal region becomes distorted, reflecting the structural asymmetry introduced by the missing atom.

Overall, the wavelike patterns observed here align well with previously reported features of the MBD model in \cite{Hauseux_NC,wavelike_science_2016}, highlighting the non-trivial many-body correlation effects captured by MBD. While a detailed physical analysis of these features lies beyond the current scope, the presented decomposition convincingly demonstrates its utility as a physically consistent and insightful analytical tool. As discussed throughout this section, our atom--atom decomposition closely mirrors a standard PW interaction approach while accurately reflecting the intrinsic geometric and physical properties of the studied molecular systems. This makes it a reliable and effective framework for analyzing MBD interactions in complex molecular systems.

\section{Towards ML surrogate modeling of MBD}
\label{sec:MLMBD}
The proposed reformulation and decomposition of MBD introduce a novel and physically meaningful representation that is well-suited for surrogate modeling. The atom-wise contribution to MBD forces is constructed as the analytical $\nabla c$ term multiplied by the many-body scaling factor $b$ which is the primary computational bottleneck. This formulation directly identifies the key quantity that a surrogate model must predict. Given the decaying nature of vdW dispersion and the resulting cutoff in dipole--dipole interactions, MBD force evaluations for a given atom can be effectively reformulated as the prediction of the corresponding $b$ terms. Compared to direct force vector prediction, learning $\boldsymbol{B}$ as an intermediate and expressive quantity provides a more physically grounded approach to capturing many-body correlations, offering a more generalizable framework for MBD surrogate modeling. 
 
\begin{figure}[htbp]
\centering
\includegraphics[trim = 20mm 50mm 20mm 50mm, clip=true,width=1\textwidth]{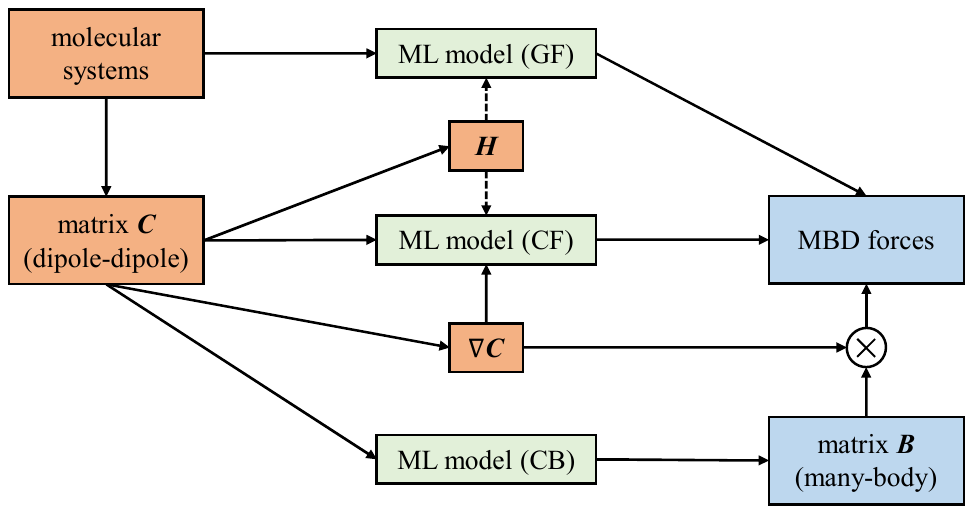}
    \caption{Roadmap for ML surrogate modeling of MBD. Three types of ML models are considered, corresponding to three different paths. Model (GF) directly predicts MBD forces from the geometric and chemical information of the molecular systems. Model (CB) instead maps matrix $\boldsymbol{C}$ to matrix $\boldsymbol{B}$. The third model (CF) takes both $\boldsymbol{C}$ and $\nabla\boldsymbol{C}$ as inputs and directly outputs MBD forces. The Hessian $\boldsymbol{H}$ could be involved in the loss function to regularize the prediction to the force.}
     \label{fig:Bmat_ML_diagram}
\end{figure}

Traditionally, MLFF models rely on geometric and chemical features of the molecular system as input and directly predict forces (see model (GF) in Fig.~\ref{fig:Bmat_ML_diagram}). As a result, their performance is highly sensitive to the encoding strategy and system complexity, often requiring architecture modifications for different molecular systems. In contrast, the proposed matrix $\boldsymbol{B}$ is derived from a non-trivial transformation of the dipole-dipole coupling matrix $\boldsymbol{C}$ that naturally serves as a descriptor of the system within the MBD formalism. This enables the surrogate model to be formulated as a matrix mapping of the characteristic $\boldsymbol{C}$ to $\boldsymbol{B}$ (model (CB) in Fig.~\ref{fig:Bmat_ML_diagram}), improving robustness while reducing the need for extensive feature encoding. Consequently, this approach provides a more general framework for surrogate modeling, as it focuses solely on learning the intrinsic relationship between $\boldsymbol{C}$ and $\boldsymbol{B}$, making it ideally applicable to arbitrary systems. 

Nevertheless, this approach does not provide a fully universal solution for MBD, as the generalization capability of any ML surrogate remains inherently constrained by the dataset. The model must still capture the underlying patterns encoded in $\boldsymbol{C}$. A truly universal surrogate for predicting $\boldsymbol{B}$ would be overly ambitious, as it is effectively equivalent to a universal solver for matrix diagonalization, a concept that belongs strictly to pure linear algebra. Moreover, despite the advantages of leveraging $b$ terms, this approach requires predicting significantly more floating-point numbers compared to the three components of a force vector for a single atom. Specifically, for a cutoff region containing $N_\text{cut}$ atoms, the model must predict $6(N_\text{cut}-1)$ values to calculate the force for the center atom, accounting for the symmetric $3\times3$ blocks that constitute non-zero entries of $\nabla\boldsymbol{C}$. Therefore, achieving highly precise force predictions becomes more challenging, as the accuracy depends on the cumulative error across all predicted terms. As a compromise, one possible solution is to give up the explicit prediction of $\boldsymbol{B}$ and instead use both $\boldsymbol{C}$ and $\nabla\boldsymbol{C}$ as inputs to an ML model, i.e., model (CF) in Fig.~\ref{fig:Bmat_ML_diagram}, which directly outputs the force vector. This approach requires an architecture capable of mimicking the methodology of MBD decomposition while retaining $\boldsymbol{B}$ in a latent representation. By doing so, the model reduces the number of values that must be predicted with high precision, thereby mitigating the accumulation of numerical errors.

The Hessian calculated according to Eq.~\eqref{eq:hessian} and Eq.~\eqref{eq:gradB_inX} provides another potential computational benefit for surrogate modeling. With an efficient implementation of the exact analytical Hessian expression, we can enrich the ML model's loss function with interatomic Hessian data involving the target atoms. While this inclusion demands higher-order differentiability from the ML architecture and increases training costs, it effectively regularizes the model and enhances generalization, analogous to the core concept behind Physics-Informed Neural Networks (PINNs) \cite{PINN}. Practically, given the rapid decay of vdW interactions, it is typically sufficient to include Hessians involving only a limited number of neighboring atoms. This localization substantially reduces the data collection burden and computational complexity, ensuring the feasibility of Hessian-enhanced training.

We acknowledge that the above discussion is prospective, and its feasibility requires further investigation. For ad hoc applications, model (GF) may offer a more efficient solution \cite{SchNet_MBD_MLST}, however, the proposed matrix-transformation-based approach provides a more generalized and physically interpretable framework for surrogate modeling. We also note that a preliminary sensitivity analysis of $\boldsymbol{B}$ under controlled perturbations of $\boldsymbol{C}$ may offer useful insight into the stability and structure of the underlying mapping, further supporting future ML development. We open-source the TensorFlow-based implementation for the data generation of necessary quantities ($\boldsymbol{C}$, $\nabla\boldsymbol{c}$, $\boldsymbol{B}$, $b\nabla c$) \cite{MBD_reform_Repo} that are potentially useful for data analysis and ML applications, while in-depth review and discussion of suitable ML architectures are left for future work.

\section{Conclusions and future work}
\label{sec:conclusion}
In this work, we introduced a structured force reformulation of the MBD model, highlighting its capability for a physically consistent and insightful decomposition of MBD forces into atom--atom contributions. Central to this reformulation is the introduction of a many-body correlation factor, represented by the matrix $\boldsymbol{B}$, which explicitly separates many-body correlation effects from the underlying pairwise dipole–-dipole coupling matrix $\boldsymbol{C}$. This new formulation provides unified and transparent expressions for the MBD energy, forces, and Hessian, significantly enhancing interpretability.

By applying our reformulation and decomposition approach to simple yet representative model systems, parallel carbon chains and rings, we demonstrated how geometric factors, particularly symmetry, strongly influence many-body correlation patterns. The force decomposition clearly revealed intriguing wavelike force profiles, providing direct insights into how MBD interactions differ fundamentally from classical pairwise methods. These characteristic patterns also serve as valuable benchmarks for future analyses of MBD interactions in complex systems.

Importantly, our formulation offers an avenue towards efficient and physically interpretable surrogate modeling of MBD interactions. The structured representation based on the matrix $\boldsymbol{B}$, combined with the atom--atom decomposition, presents a robust and expressive target for ML models. Data generation codes are provided to facilitate further development.  

\section*{Acknowledgments}
We are grateful for the support of the Luxembourg National Research Fund (C20/MS/14782078/QuaC). 

\appendix
\section{Gradient expression for \texorpdfstring{$\boldsymbol{B}$}{B}}
\label{sec:app_gradB}
Due to the commutative constraint, deriving the analytical gradient for matrix functions is more subtle compared to scalar cases. Here, we derive the gradient of the many-body matrix $\boldsymbol{B}$ explicitly in terms of the dipole coupling matrix $\boldsymbol{C}$ and its eigenvalues and eigenvectors. We begin with the relation:
\begin{equation}
    \boldsymbol{B}^2 = \boldsymbol{C}^{-1}.
    \label{eq:B2_C_1}
\end{equation}
Taking the gradient of both sides yields:
\begin{equation}
\left(\nabla\boldsymbol{B}\right)\boldsymbol{B}+\boldsymbol{B}\left(\nabla\boldsymbol{B}\right)=-\boldsymbol{C}^{-1}\left(\nabla\boldsymbol{C}\right)\boldsymbol{C}^{-1},
\label{eq:gradB^2}
\end{equation}
where the identity $\nabla\left(\boldsymbol{C}^{-1}\right)=-\boldsymbol{C}^{-1}\left(\nabla\boldsymbol{C}\right)\boldsymbol{C}^{-1}$ can be easily verified by differentiating $\boldsymbol{C}\boldsymbol{C}^{-1}=\boldsymbol{I}$. 

Now, in order to show the equivalence used in Eq.~\eqref{eq:F_mbd_reform}, we transform Eq.~\eqref{eq:gradB^2} by multiplying it by $\boldsymbol{B}^{-1}$ from the left and by $\boldsymbol{C}$ from the right, utilizing the inverse of Eq.~\eqref{eq:B2_C_1}, and applying the trace operator to both sides:
\begin{equation}
\Tr\left(\boldsymbol{B}^{-1}\left(\nabla\boldsymbol{B}\right)\boldsymbol{B}^{-1}+\left(\nabla\boldsymbol{B}\right)\boldsymbol{C}\right)=-\Tr\left(\boldsymbol{B}\left(\nabla\boldsymbol{C}\right)\right).
\end{equation}
Then, we use the fact that the trace operator allows for flexible rearrangement of matrices due to its cyclic invariance, obtaining the following relation:
\begin{equation}
    2\Tr\left(\boldsymbol{C}\left(\nabla\boldsymbol{B}\right)\right)=-
    \Tr\left(\boldsymbol{B}\left(\nabla\boldsymbol{C}\right)\right),
\end{equation}
which directly supports the MBD force reformulation presented in Eq.~\eqref{eq:F_mbd_reform}.

\emph{Remark.} It is significantly more complicated to explicitly express $\nabla\boldsymbol{B}$, as it cannot rely directly on the properties of trace operation. To do that, we first recognize that Eq.~\eqref{eq:gradB^2} takes the form of a Sylvester equation:
\begin{equation}
    \boldsymbol{\Lambda}^{-\frac{1}{2}}\boldsymbol{X}+\boldsymbol{X}\boldsymbol{\Lambda}^{-\frac{1}{2}}=-\boldsymbol{\Lambda}^{-1}\boldsymbol{Y}\boldsymbol{\Lambda}^{-1}
\end{equation}
where $\boldsymbol{X}=\boldsymbol{S}^{\text{T}}\left(\nabla\boldsymbol{B}\right)\boldsymbol{S}$ and $\boldsymbol{Y}=\boldsymbol{S}^{\text{T}}\left(\nabla\boldsymbol{C}\right)\boldsymbol{S}$, while $\boldsymbol{\Lambda}$ and $\boldsymbol{S}$ represent the eigen system of $\boldsymbol{C}$. The standard method for solving the Sylvester equation is the Bartels–Stewart algorithm \cite{BS-algorithm}, which requires an $O(N^3)$ computational cost. However, in our case, the diagonal structure of the eigenvalue matrix $\boldsymbol{\Lambda}$ simplifies the problem, allowing for a direct component-wise expression for~$\boldsymbol{X}$:
\begin{equation}
\boldsymbol{X}_{pq} = -\frac{\boldsymbol{Y}_{pq}}{(\lambda_p^{\frac{1}{2}} + \lambda_q^{\frac{1}{2}}) \lambda_p \lambda_q},
\label{eq:gradB_inX}
\end{equation}
where $\lambda_p$ and $\lambda_q$ are the diagonal elements of $\boldsymbol{\Lambda}$ and correspond to the eigenvalues of the matrix $\boldsymbol{C}$. Finally, we transform $\boldsymbol{X}$ back to the original basis to recover the gradient as: $\nabla\boldsymbol{B}= \boldsymbol{S}\boldsymbol{X}\boldsymbol{S}^{\text{T}}$. 

 \bibliographystyle{elsarticle-num}
 \bibliography{cas-refs}






\end{document}